*Original Article*

## Effect of Copper Chemical Form on The Growth of *Pseudomonas aeruginosa* Isolated from Burned Patients and on Its Cu Uptake

Amjad Al Tarawneh[a], Haitham Qaralleh[b], Muhamad O. Al-limoun[c], Khaled M. Khleifat[c]

[a]Prince Faisal Center for Dead Sea, Environmental and Energy Research (PFC-DSEER), Mutah University, Karak, Jordan, Mutah 61710, P.O.Box: 7; [b]Department of Medical Laboratory sciences, Mutah University, Karak, Jordan; [c]Department of Biology, Mutah University, Karak, Jordan

*Corresponding Author: amjtar@hotmail.com, amjtar@mutah.edu.jo

**Abstract**: The main object of this investigation was to shed light on the information on the level of Cu uptake by *Pseudomonas aeruginosa* that has been previously isolated from patients using Gram-negative *E. aerogenes* ans a Gram-positive *Bacillus thuringiensis* as a control for Cu uptake measurements. Cu uptake data showed that maximum Cu uptake was obtained by the *Pseudomonas aeruginosa*. The metal uptake was dependent on the type of biosorbent with different accumulation affinities toward the tested chemical form of the copper. Cells harvested at exponential growth phase showed slightly higher Cu uptake than at stationary phase, which reflect that the Cu uptake is metabolism dependent-process. The increasing order of affinity of the cupric chloride and cupric sulphate towards the three genera were almost constant. However, where cupric nitrate was used, the copper uptake behavior in Pseudomonas aeroginosa was not changed from those results with the other two chemical forms, whereas with *E. aerogenes* & *B. thuringiensis*, different Cu uptake behaviors were observed. In *B. thuengenesis*, the copper uptake rate as a function of initial Cu concentration were shown to increase with increasing the initial concentration of Cu, constantly above 320 ppm as compared with others. In *Enterobacter* with cupric nitrate, different bahavior was also shown in which the copper uptake decline sharply beyond the 320 ppm. These results reflect that the Cu uptake in *P. aeruginosa* is with different mechanisms as compared with other tested bacteria.

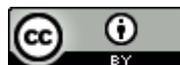

## INTRODUCTION

Pseudomonas aeruginosa cause high percentage of opportunistic infections. Pseudomonas aeruginosa infection is a significant issue in patients hospitalized with burns along with other diseases like cancer, cystic fibrosis; which causes a 50% fatality of them. Different infections include pneumonia, endocarditis, and infections of the wounds, urinary tract, ears, central nervous system, musculoskeletal system, eyes, skin, and also could be caused by Pseudomonas species (Bodey et al., 1993; Botzenhart and Doring, 1933). The bacterium Pseudomonas contains tens of species probly upto 150, most of which are saprophytic. More than two dozens of species are linked with humans. Generally, pseudomonads are well-known to lead to ailments in humans are correlating to the opportunistic infections (McManus et al., 1985; Shankowsky et al., 1994). This bacterial infection lead to common complications which participate substantially to burn morbidity and mortality. The copper is an essential trace element for living organisms including bacteria, which takes part with the iron in a unique ability to maintain special redox states for biological functions carrier proteins as well as their need for growth and development (Linder et al., 1999; Pena et al., 1999; Khleifat, 2006; Khleifat et al., 2006a). Many agents have an crucial impact on the microbial growth including and metals uptake (i.e., temperature) or that pertain to the natural characteristics of the bacteria itself. Studies have been made that produced various kinetic models for predicting the growth mode of microorganism, especially lag time and maximum specific growth rate. Also, copper ion can be potent cytotoxin when accumulated in excess of cellular requirement (Linder and Hazegh-Azam, 1996; Lu and Solioz, 2001). Therefore, the maintenance of appropriate Cu homeostasis requires certain balance between its cellular uptake and detoxification (Lu and Solioz, 2001).

In bacteria, common resistance mechanisms including the efflux systems for managing the heavy metals such as copper. One of these mechanisms is the *cop* system that containing the structural genes *cop ABCD* (Cooksey, 1994; Silver, 1996). The *cop B* and *cop D* genes are involved in the transport of copper across the membrane, while the products of *cop A* and *cop C* genes are outer membrane proteins that bind Cu in the periplasm to protect the cell from copper toxicity. Other types of efflux systems such as the P-type ATPase, simply operate through pumping toxic Cu ions out of the cell .Interestingly, the P-type ATPases defective in human hereditary diseases of copper metabolism were found to have more similarity to bacterial than other eukaryotes ATPases (Silver, 1996).

Numerous pseudomonas strains have availability for *in vitro* copper uptake (Hussein et al., 2005; Chen et al., 2006). The strain *Pseudomonas aeruginosa* particularly, has unique ability to resist high levels of some xenobiotics like antimicrobial agents, solvents, and heavy metals (Wang et al., 1997; Khleifat, 2010; Khleifat et al., 2019).This resistance is attributed to a combination of decreased outer membrane permeability and the presence of multiple efflux pumps (Nikadio, 1996; Nikadio, 1997). The present work was carried out to parallel the efficacy of *in vitro* copper uptake by *Pseudomonas aeruginosa* and control bacterial strains. In addition, the object of this investigation was to shed light on the information on the level of Cu uptake by Pseudomonas aeroginosa and its growth on different chemical forms of copper compounds.

## MATERIALS AND METHODS
### Bacterial isolation and growth
The strains *Pseudomonas aeruginosa*, *Pseudomonas putrefaciens, B. thuringiensis* and *E. aerogenes* were grown on Luria-Bertani (LB) medium (Khleifat and Abboud, 2003) .This medium is composed of 10 g trypton, 5 g of yeast extract and 10 g of NaCl per litter of distilled water. The pH was adjusted to 7.0 with 0.1 N NaOH or 0.1 N HCl. The LB agar plates containing 15 g/L of agar in addition to the above mixture. An initial inoculum of $1.5 \times 10^8$ viable cells was used in all cultures. The growth was maintained in 50 ml LB-broth media using agitation rate of 150 rpm and 37C°. The pathogenic strain *Pseudomonas aeruginosa* was isolated from the blood and scar tissues of hospitalized burn patients .The isolated strain was characterized morphologically and analyzed by the Api 20 NE Kit testing system (Biomeriex, France)as described before (Dance et al., 1989; Khleifat and Abboud, 2003; Abboud et al., 2010; Khleifat et al., 2014; Khleifat et al., 2015; Majali et al., 2015; Althunibat et al., 2016).

### Burn patients
Forty two patients (males and females ranged between 19-42 years of age) were admitted to hospital at different intervals suffering second or third degree of burning. On first day of admission, blood samples were collected from these patients and considered as pre-infected burn control. Five days after hospitalization, these burn patients usually become more susceptible for infection with *Pseudomonas aeruginosa* .The infection with this bacterial strain was confirmed later by conducting the above mentioned Api 20 NE testing system on corresponding samples from burn patients.

### *In vitro* bacterial copper uptake
Copper solutions of either cupric chloride dehydrate or cupric nitrate 3 hydrate or cupric sulfate were prepared in dis. water at different concentrations range of 20, 40, 60, 80, 100, 200, 300, 400, 500, 600, and 700 part per million (ppm), respectively. These solutions were sterilized in the autoclave. Meanwhile, Fifty mg of bacterial biomass was harvested from exponentially growing bacteria in LB medium and washed twice with 5 ml Ringer's solution (NaCl 0.85%, $CaCl_2$ 0.03%, KCl 0.025%, $NaHCO_3$ 0.02%) using refrigerated centrifuge.

To perform an *in vitro* copper uptake experiment, the method described before (Vivas et al., 2006) was applied with slight modifications. The harvested bacterial biomass was incubated for 1 hr with the corresponding copper compound concentration at 37C° and 150 rpm agitation rate. At the end of incubation period the biomass was harvested and washed with 0.1M ammonium acetate solution. The washed biomass was decomposed by 1% nitric acid for 24 hr and the biomass copper contents were quantified by atomic absorption spectroscopy.

### Data analysis
The results were expressed as mean ± SD and analyzed statistically by student's t-test.
The correlation between the data was tested by simple linear regression, employing SPSS computer program. P value of less than 0.05 was considered as the lowest limit of significance (Tarawneh et al., 2011).

## RESULTS
### Biomass
The effect of different cupric compounds and concentrations for these variable chemical forms of the copper were studied using cell biomass as parameter. Table 1 show the results of *Pseudomonas aeruginosa* that was grown on LB-media at 37°C as mentioned in Materials and Methods. The biomass g/L was differently affected by the copper concentration and it's chemical forms. Two copper compounds (cupric acetate monohydrate and cupric sulfate anhydrus) gave the lowest biomass and were excluded from the forther experiments, however cupric chloride dihydrate, cupric nitrate 3 hydrate and cupric sulfate at concentration of 300 ppm resulted in highest cell yield (g/L). The results are also shown in (Figure 1) in which beyond the concentration of 300 ppm the inhibition was progressively as concentration increased. Whereas cupric sulfate at 500 ppm the yield was the highest (14.86 g/L) and the cupric acetate monohydrate the highest biomass at 1000 ppm.

Upon the results obtained from (Table 1), two concentrations of copper compounds were choosed, 300 and 1000 ppm to study the growth curves by viable account. Also wide range of copper compounds concentration were used (20, 40, 60, 80, 100, 200, 300, 400, 500, 600 and 700 ppm) for the measurement of copper uptake, biosorption and adsorption, mostly from these concentrations showed exponentially increase in the biomass of the (Figure 1)

**Growth curves**

As mentioned previously only three compounds (cupric chloride dihydrate, cupric nitrate 3 hydrate and cupric sulfate) with concentration (300 and 1000 ppm) were used in studding the profile of growth curves.The viable cells were precisely enable us to see the effect of copper compounds on the growth kinetics of the bacteria *Pseudomonas aeruginosa*. The bioaviability of heavy metals in the broth is more than in the plates (Khleifat and Homady, 2000). Generally the results here were in agreement with the results of biomass. It is notable that the lag phase of this bacteria even at 0 ppm is extended as compared with other gram negative bacteria grown on LB-media (Khleifat and abboud, 2003). The growth profile for this bacteria at 0 ppm resulted in the extended lag phase of about 10 hours, however with cupric chloride dihydrate and cupric sulfate at 300 ppm resulted in more extended lag phase at about 12 hours and 14 hours for cupric nitrate 3 hydrate. The existence of heavy metals in the growth media probably causing shock to the bacteria and even after hours the bacteria make adaptation and start growing (Khleifat et al., 2006b).

The growth of *Pseudomonas aeruginosa* at 300 ppm of the three compounds resulted in the almost 65% inhibition as compared with the growth at 0 ppm (Figure 2). Also the bacteria at 0 ppm have been shown to grow faster than the same bacteria at 300 ppm cupric compounds. However this bacterium at the same concentration of 300 ppm for the cupric compounds were differently showed growth kinetics, For example, with cupric chloride dihydrate and cupric sulfate were identical, whereas, with the cupric nitrate 3 hydrate resulted in the slowest growth pattern as compared with that obtained with other chemical forms. In the meanwhile the growth profile with cupric nitrate showed earlier inhibition as compared with the growth of *Paeudomonas aeruginosa* with other two cupric compounds. Finally the concentration of 1000 ppm of the three chemical forms of the copper wee shown to be lethal (Figure 2).

**Copper uptake.**

As mentioned in Materials and Methods, three bacterial genera were used to measure the copper uptake. These genera include the major genus in our study, the *Pseudomonas aeruginosa* and the other two controls one of them is gram negative (*E. aerogenes*) and the second control is a gram positive bacteria (*B. thurengenesis*. Strain *israelis*). In this study the amount of copper enter the cells was measured according to the method of (Kanazawa and Mor, 1996) and reflect to as uptake. Using the cupric chloride dihydrate as copper cation to measure it's uptake by the three mentioned bacteria was shown to be differently affected the level of copper uptake (data not shown). In *Pseudomonas aeruginosa*, the high copper uptake was shown to be at 60 ppm copper concentration and resulted in the amount of 470.48 ppm/g biomass. In *Bacillus thurengenesis*, the maximum copper uptake was at 100 ppm copper concentration and equal to 350.292 ppm/g biomass. In *E. aerogenese* the maximal copper uptake is at 200 ppm copper concentration and resulted in 383.356 ppm/g biomass. The data here, show that copper uptake by the three bacteria is comparisonable. The copper uptake by the main sample bacteria *Pseudomonas aeruginosa* is 25% higher than the copper uptake by *Bacillus thurengenesis* and 18% higher than that obtained by *E. aerogenes* (Table 2).

**Table 1.** Effect of copper compounds on the biomass of *Pseudomonas aeruginosa* g/L

| concentration | cupric chloride dihydrate | cupric nitrate 3 hydrate | cupric sulphate anhydrous | cupric acetate monohydrate | cupric sulphate |
|---|---|---|---|---|---|
| 100 ppm | 11.86 | 9.78 | 9.86 | 10.64 | 10.24 |
| 300 ppm | 16.12 | 13.78 | 11.6 | 4.44 | 13.76 |
| 500 ppm | 5.04 | 11.66 | 4.84 | 5.18 | 14.86 |
| 700 ppm | 4.18 | 3.8 | 4.24 | 4.24 | 3.84 |
| 1000 ppm | 5.34 | 4.84 | 4.18 | 4.42 | 4.06 |

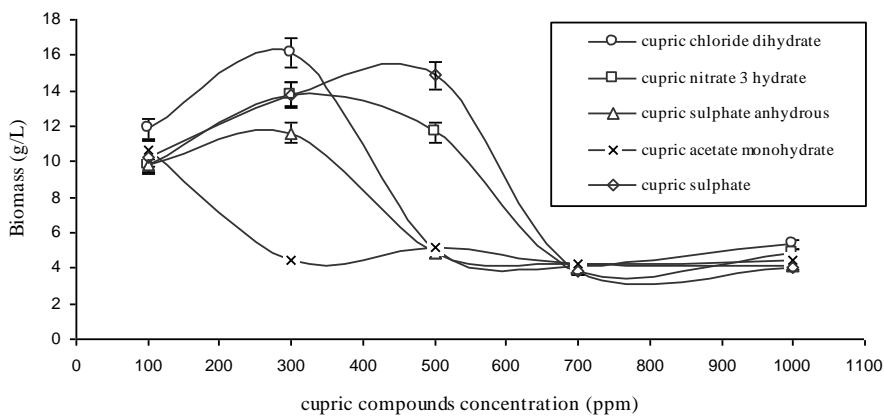

Figure 1. Biomass g/L of *Pseudomonas aeruginosa* cultured in five copper compounds with different concentrations

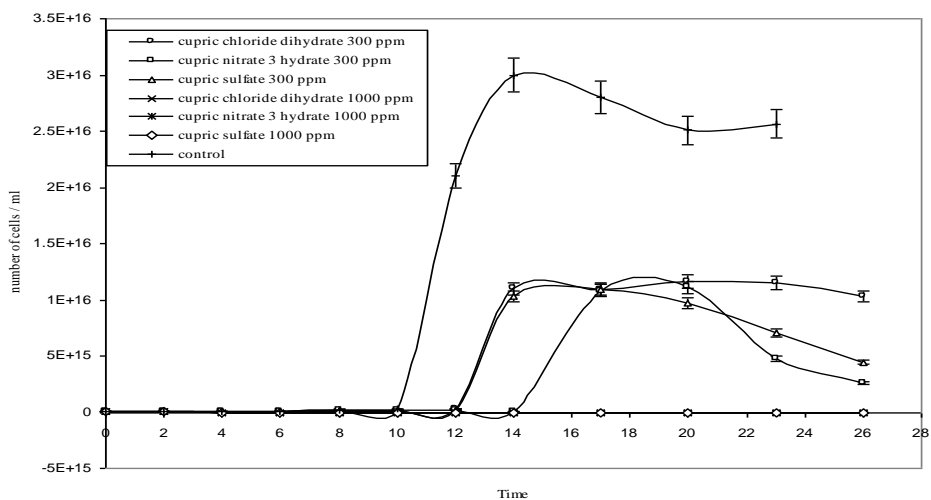

Figure 2. Growth curve for *Pseudomonas aeruginosa* at (300 ppm and 1000 ppm) concentrations of three copper compounds, compared with growth of *Pseudomonas aeruginosa* at 0 ppm (control).

The copper uptake results of three bacteria with this compound reflects different affinities to this heavy metals and consequently indicates why the patients infected with *Pseudomonas aeruginosa* suffered of copper deficiency (Abboud et al., 2009). Using the cupric sulfate as a substrate to measure the copper uptake by the same genera. The results here, were parallel with that obtained when using the cupric chloride dihydrate. However, the copper uptake by the three bacteria were slightly lower than of that obtained by cupric chloride dihydrate but with the same varieties among the three bacteria (Table 3).

In the contrary using the cupric nitrate 3 hydrate with three bacteria (Table 4), the copper uptake results show different pattern of that seen by cupric chloride dihydrate and cupric sulfate. The copper uptake by *Pseudomonas aeruginosa* almost identical with that obtained from using cupric chloride dihydrate (Table 5). *E. aerogenese* also show the same results and intercept at 320 ppm copper concentration, with that of *Bacillus thurengenesis*. Beyond the 320 ppm the copper uptake is decrease with increasing the concentration. In *Bacillus thurengenesis*, the copper uptake is total different from the previous data in which the copper uptake is progressively increase as the copper concentration increased. A result had not been seen with other compounds and other two genera used, It is evidence that the copper uptake here in *Bacillus thurengenesis* is with different mechanisms.

Table 2. Effect of cupric chloride dihydrate on the copper uptake ppm/g wet weight in three bacterial strains, *P. aeruginosa B. thuringiensis* and *E. aerogenes*

| Cupricchloride dihydrate concentration (ppm) | Copper Uptake ppm/g wet weight (*P. aeruginosa*) | Copper Uptake ppm/g wet weight (*B. thuringiensis*) | Copper Uptake ppm/g wet weight (*E. aerogenes*) |
|---|---|---|---|
| 20 | 376.36 | 269.62 | 298.72 |
| 40 | 456.68 | 280.66 | 321.04 |
| 60 | 470.48 | 338.6 | 346.6 |
| 80 | 472.52 | 345.08 | 350.82 |
| 100 | 455.428 | 350.292 | 365.392 |
| 200 | 438.814 | 326.808 | 383.356 |
| 300 | 436.744 | 319.462 | 380 |
| 400 | 468.05 | 289.878 | 395.72 |
| 500 | 464.47 | 316.07 | 392.15 |
| 600 | 490.026 | 261.392 | 355.986 |
| 700 | 501.926 | 239.396 | 355.006 |

**Table 3.** Effect of cupric sulfate on the copper uptake ppm/g in three bacterial strains, *Pseudomonas aeruginsa*, *Bacillus thurengenesis* and *E. aerogenese*

| Cupric sulfate concentration (ppm) | Copper Uptake ppm/g wet weight (*P. aeruginosa*) | Copper Uptake ppm/g wet weight (*B. thuringiensis*) | Copper Uptake ppm/g wet weight (*E. aerogenes*) |
|---|---|---|---|
| 20 | 152.84 | 121.82 | 84.3 |
| 40 | 384.18 | 273.62 | 293.28 |
| 60 | 395.12 | 290.18 | 295.82 |
| 80 | 401.23 | 285.08 | 310.88 |
| 100 | 406.718 | 279.351 | 297.18 |
| 200 | 396.252 | 253.176 | 298.18 |
| 300 | 386.854 | 277.988 | 340.55 |
| 400 | 390.58 | 277.618 | 355.65 |
| 500 | 365.72 | 294.89 | 356.104 |
| 600 | 369.314 | 272.304 | 346.592 |
| 700 | 381.878 | 301.332 | 256.986 |

Table 4. Effect of cupric nitrate 3 hydrate on the copper uptake ppm/g in three bacterial strains, *Pseudomonas aeruginosa*, *Bacillus thurengenesis* and *E. aerogenese*

| Cupric nitrate 3 hydrate concentration (ppm) | Copper Uptake ppm/g wet weight (*P. aeruginosa*) | Copper Uptake ppm/g wet weight (*B. thuringiensis*) | Copper Uptake ppm/g wet weight (*E. aerogenes*) |
|---|---|---|---|
| 20 | 344.6 | 98.28 | 187.36 |
| 40 | 439.78 | 112.5 | 328.5 |
| 60 | 450.27 | 140.18 | 337.34 |
| 80 | 440.96 | 156.7 | 349.44 |
| 100 | 445.96 | 200.14 | 341.346 |
| 200 | 440.53 | 235.778 | 300.536 |
| 300 | 453.64 | 287.948 | 295.844 |
| 400 | 444.624 | 299.024 | 266.482 |
| 500 | 471.708 | 349.77 | 244.668 |
| 600 | 475.932 | 344.246 | 195.022 |
| 700 | 413.564 | 345.385 | 151.046 |

Table 5. Effect of three copper compounds (cupric nitrate 3 hydrate, cupric chloride dihydrate and cupric sulfate) on the copper uptake ppm/g in *Pseudomonas aeruginosa.*.

| copper concentration (ppm) | Copper Uptake ppm/g wet weight of cells grown with cupric nitrate 3 hydrate | Copper Uptake ppm/g wet weight of *cells* grown with cupric chloride dihydrate | Copper Uptake ppm/g wet weight of cells grown with cupric sulfate |
|---|---|---|---|
| 20  | 344.6   | 376.36  | 152.84  |
| 40  | 439.78  | 456.68  | 384.18  |
| 60  | 450.27  | 470.48  | 395.12  |
| 80  | 440.96  | 472.52  | 401.23  |
| 100 | 445.96  | 455.428 | 406.718 |
| 200 | 440.53  | 438.814 | 396.252 |
| 300 | 453.64  | 436.744 | 386.854 |
| 400 | 444.624 | 468.05  | 390.58  |
| 500 | 471.708 | 464.47  | 365.72  |
| 600 | 475.932 | 490.026 | 369.314 |
| 700 | 413.564 | 501.926 | 381.878 |

**DISCUSSION**

Present data indicate superior abilities by P. *aeruginosa* for the uptake as well as tolerance of copper ions than a selected control of two non-pathogenic Gram positive and Gram negative bacterial strains. Moreover, the *Pseudomonas aeruginosa* strain showed much higher efficiency for copper uptake than the weakly pathogenic strain *Pseudomonas (Shewanella) putrefaciens*. The later strain is a saprophytic gram-negative rod, which is rarely implicated in severe clinical syndromes of human (Chen et al., 1997; Khase and Jande, 1998; Pagani et al., 2003). Copper removal from the environment by *P. aeruginosa* is usually carried out by an outer membrane protein namely protein C (OprC) that functions as channel-forming and copper binding protein (Hall et al., 2001) .As obligate aerobe, *Pseudomonas aeruginosa* possesses stringent requirement for metal ions like iron (Cox, 1989) and possibly for copper to facilitate its respiratory enzymes mediated aerobic metabolism during multiplication in the infected host (Ochsner et al., 2000)**.**

Concomitantly with the efficiency of copper uptake, a particular hypocupremia in burn patients infected by *P. aeruginosa* was shown to be absent in matched group of burn patients free of infection (Abboud et al., 2009). Similarly a decrease in copper contents of the plasma was linked with the operation of an effective efflux mechanism for the transport of copper by Plasmodium that infects erythrocytes (Rasoloson et al., 2004). Energetically growing bacteria experienced to an elevated concentration of copper for a stumpy period of schedule was acknowledged as a "copper-shocked" bacterial culture. On the contrary, the populations that were adapted to copper perceived as cells effectively growing while copper is elevated. A 405 genes expression was changed in the copper-shocked culture, whereas only 331 genes were expressed in the copper-adapted bacterial cultures. This in turn, was indicated by identified common genes for both conditions. For instance, the two stress conditions lead to up-regulation of genes encoding several active transport functions. However, there were some fascinating variations between the two types of stress. Only the cells adapted to copper were significantly altered the expression of passive transport functions and several porins belonging to the OprD family down-regulated their expression. Copper shock lead to the expression profiles revealing to the response of oxidative stress , maybe as a result of copper participation in Fenton-like chemistry. However, cells that was adapted to copper had no such a response (Yoneyama and Nakae, 1996; Teitzel et al., 2006). It was reported that, microorganisms are able to adjust to the existence of toxic substances by using a succession of adaptation strategies such as alteration in lipid composition of cell membranes to recompense for these substances induced surge in membrane fluidity (Khleifat, 2007; Khleifat et al., 2008).

Bacterial resistance to copper metal has demonstrated an important contribution to the competitive endurance of some pseudomonads in such environment (Yang et al., 1993). specially during infection, resistance to heavy metals by P. aeruginosa may be a health concern as a result of its ability to cause additional resistance to antibiotics. Perron et a.l (2004) found that P. aeruginosa strains which were picked up on zinc became resistant to both heavy metals (zinc, cadmium, and cobalt) and to the carbapenem antibiotic imipenem (Abboud et al., 2009). The selection of cross-resistance selection was attributed to a common co-regulation mechanism between carbapenem influx and heavy metal efflux .The investigation of analogues cross-resistance mechanism between the copper efflux and antibiotics influx in burn patients infected with P. aeruginosa might be extra in need to pursue it in the days to come.